\documentstyle[12pt]{article}

\catcode`\@=11
\long\def\@makefntext#1{
\protect\noindent \hbox to 3.2pt {\hskip-.9pt
$^{{\ninerm\@thefnmark}}$\hfil}#1\hfill}		

\def\@makefnmark{\hbox to 0pt{$^{\@thefnmark}$\hss}}  

\def\ps@myheadings{\let\@mkboth\@gobbletwo
\def\@oddhead{\hbox{}
\rightmark\hfil\ninerm\thepage}
\def\@oddfoot{}\def\@evenhead{\ninerm\thepage\hfil
\leftmark\hbox{}}\def\@evenfoot{}
\def\sectionmark##1{}\def\subsectionmark##1{}}

\renewcommand{\thefootnote}{\fnsymbol{footnote}}

\newcounter{sectionc}\newcounter{subsectionc}\newcounter{subsubsectionc}
\renewcommand{\section}[1] {\vspace*{0.6cm}\addtocounter{sectionc}{1}
\setcounter{subsectionc}{0}\setcounter{subsubsectionc}{0}\noindent
	{\normalsize\bf\thesectionc. #1}\par\vspace*{0.4cm}}
\renewcommand{\subsection}[1] {\vspace*{0.6cm}\addtocounter{subsectionc}{1}
	\setcounter{subsubsectionc}{0}\noindent
	{\normalsize\it\thesectionc.\thesubsectionc. #1}\par\vspace*{0.4cm}}
\renewcommand{\subsubsection}[1]
{\vspace*{0.6cm}\addtocounter{subsubsectionc}{1}
	\noindent {\normalsize\rm\thesectionc.\thesubsectionc.\thesubsubsectionc.
	#1}\par\vspace*{0.4cm}}

\newcounter{appendixc}
\newcounter{subappendixc}[appendixc]
\newcounter{subsubappendixc}[subappendixc]

\renewcommand{\appendix}[1] {\vspace*{0.6cm}
        \refstepcounter{appendixc}
        \setcounter{figure}{0}
        \setcounter{table}{0}
        \setcounter{equation}{0}
        \renewcommand{\thefigure}{\Alph{appendixc}.\arabic{figure}}
        \renewcommand{\thetable}{\Alph{appendixc}.\arabic{table}}
        \renewcommand{\theappendixc}{\Alph{appendixc}}
        \renewcommand{\theequation}{\Alph{appendixc}.\arabic{equation}}
        \noindent{\bf Appendix \theappendixc #1}\par\vspace*{0.4cm}}

\def\abstracts#1{{

\centering{\begin{minipage}{12.2truecm}\footnotesize\baselineskip=12pt\noindent
	\centerline{\footnotesize ABSTRACT}\vspace*{0.3cm}
	\parindent=0pt #1
	\end{minipage}}\par}}

\renewenvironment{thebibliography}[1]
	{\begin{list}{\arabic{enumi}.}
	{\usecounter{enumi}\setlength{\parsep}{0pt}
\setlength{\leftmargin 1.25cm}{\rightmargin 0pt}
	 \setlength{\itemsep}{0pt} \settowidth
	{\labelwidth}{#1.}\sloppy}}{\end{list}}

\newcounter{itemlistc}
\newcounter{romanlistc}
\newcounter{alphlistc}
\newcounter{arabiclistc}

\newcommand{\fcaption}[1]{
        \refstepcounter{figure}
        \setbox\@tempboxa = \hbox{\footnotesize Fig.~\thefigure. #1}
        \ifdim \wd\@tempboxa > 6in
           {\begin{center}
        \parbox{6in}{\footnotesize\baselineskip=12pt Fig.~\thefigure. #1}
            \end{center}}
        \else
             {\begin{center}
             {\footnotesize Fig.~\thefigure. #1}
              \end{center}}
        \fi}

\newcommand{\tcaption}[1]{
        \refstepcounter{table}
        \setbox\@tempboxa = \hbox{\footnotesize Table~\thetable. #1}
        \ifdim \wd\@tempboxa > 6in
           {\begin{center}
        \parbox{6in}{\footnotesize\baselineskip=12pt Table~\thetable. #1}
            \end{center}}
        \else
             {\begin{center}
             {\footnotesize Table~\thetable. #1}
              \end{center}}
        \fi}

\def\@citex[#1]#2{\if@filesw\immediate\write\@auxout
	{\string\citation{#2}}\fi
\def\@citea{}\@cite{\@for\@citeb:=#2\do
	{\@citea\def\@citea{,}\@ifundefined
	{b@\@citeb}{{\bf ?}\@warning
	{Citation `\@citeb' on page \thepage \space undefined}}
	{\csname b@\@citeb\endcsname}}}{#1}}

\newif\if@cghi
\def\cite{\@cghitrue\@ifnextchar [{\@tempswatrue
	\@citex}{\@tempswafalse\@citex[]}}
\def\citelow{\@cghifalse\@ifnextchar [{\@tempswatrue
	\@citex}{\@tempswafalse\@citex[]}}
\def\@cite#1#2{{$\null^{#1}$\if@tempswa\typeout
	{IJCGA warning: optional citation argument
	ignored: `#2'} \fi}}

 1
 1
 1

\font\ninerm=cmr9

\begin{document}

\newcommand{\st}{\scriptstyle}
\newcommand{\sst}{\scriptscriptstyle}
\newcommand{\mco}{\multicolumn}
\newcommand{\epp}{\epsilon^{\prime}}
\newcommand{\vep}{\varepsilon}
\newcommand{\ra}{\rightarrow}
\newcommand{\ppg}{\pi^+\pi^-\gamma}
\newcommand{\vp}{{\bf p}}
\newcommand{\ko}{K^0}
\newcommand{\kb}{\bar{K^0}}
\newcommand{\al}{\alpha}
\newcommand{\ab}{\bar{\alpha}}
\def\be{\begin{equation}}
\def\ee{\end{equation}}
\def\bea{\begin{eqnarray}}
\def\eea{\end{eqnarray}}
\def\CPbar{\hbox{{\rm CP}\hskip-1.80em{/}}}

\centerline{\normalsize\bf CP VIOLATION IN THE WEINBERG}
\baselineskip=22pt
\centerline{\normalsize\bf MULTI-HIGGS MODEL
\footnote{ To appear in the proceedings of the conference " Beyond the
Standard Model IV " Granlibakken, Lake Tahoe, California, Dec. 13-18 1994.
Hosted by the Davis Institute for High Energy Physics. Edited by J. Gunion,
T. Han, and J. Ohnemus.  }}
\baselineskip=16pt

\centerline{\footnotesize JOANNIS PAPAVASSILIOU}
\baselineskip=13pt
\centerline{\footnotesize\it Department of Physics, New York University,
4 Washington Place}
\baselineskip=12pt
\centerline{\footnotesize\it NY, NY 10003, USA}
\centerline{\footnotesize E-mail: papavass@mafalda.physics.nyu.edu}
\vspace*{0.3cm}

\vspace*{0.9cm}
\abstracts{We report new results in the study of CP violation in semileptonic
top decays, in the context of the Weinberg Model.}

\normalsize\baselineskip=15pt
\setcounter{footnote}{0}
\renewcommand{\thefootnote}{\alph{footnote}}
\section{Top decays in the Weinberg Model}

Semileptonic top decays in the context of the Weinberg Model
(WM) \cite{Wein} have been the focal point of extensive
study \cite{Gunion}.
In the WM the new basic ingredient is the possibility
of inducing CP violating effects in the leptonic sector, due to the presence
of the additional charged Higgs sector. The way such CP violating effects
arise can be seen from the relevant Lagrangian term
\begin{equation}
{\cal L}= \frac{gm_{t}}{\sqrt{2} M}{\bar{t}}_{R}b_{L}
\frac{c_{1}c_{2}s_{3}-s_{2}c_{3}e^{i\delta}}{s_{1}c_{2}}H^{+}
-\frac{gm_{\tau}}{\sqrt{2} M}{\bar{\nu}}_{L}\tau_{R}
\frac{c_{1}s_{2}s_{3}+c_{2}c_{3}e^{i\delta}}{s_{1}c_{2}}H^{+}
+h.c.~~,
\label{Lagr}
\end{equation}
involving Yukawa couplings between the extra charged Higgs $H^{+}$ and the
fermions (quarks and leptons).
We note that the constants $s_{i}$, $c_{i}$ and $\delta$
appear in the CKM-like matrix operating in the charged Higgs sector and are
not elements of the usual CKM matrix; $M$ is the mass of the $W$.
The possibility for additional CP violating effects has been studied in the
context of the decay mode $t\rightarrow b\tau\nu$.
The observable considered is the partial decay rate
asymmetry (PRA), namely
\begin{equation}
{\cal A}=\frac{\Gamma (t\rightarrow b \tau^{+}\nu_{\tau})
-\Gamma (\bar{t}\rightarrow \bar{b}\tau^{-}\bar{\nu_{\tau}})}
{\Gamma (t\rightarrow b \tau^{+}\nu_{\tau})
+\Gamma (\bar{t}\rightarrow \bar{b}\tau^{-}\bar{\nu_{\tau}})}~.
\label{DefA}
\end{equation}

At one loop the PRA receives contributions through interference terms
between one-loop Standard Model (SM) graphs for the process
$t\rightarrow W^{+}b\rightarrow b\tau^{+}\nu_{\tau}$,
and the tree-level WM graph for the
process $t\rightarrow H^{+}b\rightarrow b\tau^{+}\nu_{\tau}$.
Consequently, the entire effect is proportional to $m_{t}m_{\tau}$.
Due to helicity mismatches \cite{Soni1} only the
longitudinal parts of the SM graphs contribute to the PRA. In addition,
due to the fact that the Higgs couplings are complex numbers, it is only the
imaginary parts of such longitudinal contributions which is relevant.
So, ${\cal A}$ is proportional to
\begin{equation}
{\cal A}\sim \int dq^{2}f(q^{2})Im(G_{L})~,
\label{Phase}
\end{equation}
where $f(q^{2})$ is a phase space function \cite{Soni2}
and $G_{L}$ is the longitudinal component of any one-loop graph.
In addition, it is important to notice the presence of the phase space
integral, whose range extends from $m_{\tau}^{2}$ all the way up to
${(m_{t}-m_{b})}^{2}$.

In computing the one-loop contribution to ${\cal A}$ the only graphs
considered was the $W$ self-energy graphs, containing fermionic loops (
although, as we will see in the next section, they are not the
only graphs contributing to ${\cal A}$). The original motivation
for singling
out the $W$ propagator with fermionic loops was the
expectation that due to the
general {\sl resonant} nature of such graphs, significant enhancement of the
PRA might take place. As it was soon realized however \cite{Soni1}
this
resonant behavior could not
be exploited, because it is only the longitudinal parts of
the self-energy graphs which contribute to the PRA,
and
it is only the transverse (but not the longitudinal) parts of the $W$
self-energy, which displays resonant behavior. So, when the
$W$ propagator is
decomposed in the form
\begin{equation}
{G}_{\mu\nu}=(g_{\mu\nu}-\frac{q_{\mu}q_{\nu}}{q^{2}})G_{T}+
\frac{q_{\mu}q_{\nu}}{q^{2}}G_{L}~,
\label{Decomp}
\end{equation}
with
\begin{equation}
G_{T}= \frac{1}{q^{2}-M^{2}+i\epsilon_{T}}~,
\label{Pt}
\end{equation}
and
\begin{equation}
G_{L}= \frac{1}{M^{2}+i\epsilon_{L}}~,
\label{Pl}
\end{equation}
where
\begin{equation}
\epsilon_{T}=(\frac{g^{2}}{32\pi})
\frac{(2q^{2}+m_{c}^{2}){(q^{2}-m_{c}^{2})}^{2}}{q^{4}}~,
\label{Et}
\end{equation}
and
\begin{equation}
\epsilon_{L}= (\frac{3g^{2}}{32\pi})
\frac{m_{c}^{2}{(q^{2}-m_{c}^{2})}^{2}}{q^{4}}~,
\label{El}
\end{equation}
from Eq(\ref{Pt}) and Eq(\ref{Pl}) follows that
\begin{equation}
Im(G_{T})= -\epsilon_{T}{|G_{T}|}^{2},~~~
Im(G_{L})= -\epsilon_{L}{|G_{L}|}^{2}~.
\label{ImPt}
\end{equation}
We notice that
due to rescattering the
$\tau\nu$ loop should not contribute
for CPT to be an exact symmetry, so that the next
threshold is due to the $cs$ loop. Finally, when $ImG_{L}$ of
Eq(\ref{ImPt}) is
inserted in Eq(\ref{Phase}) (instead of the resonant $ImG_{T}$ which does not
contribute), the result is very small (${\cal A}\sim 10^{-8}$).

In an attempt
to exploit the resonant character of $ImG_{T}$,
one then proceeded
to compute {\sl two loop} contributions \cite{Soni1} to ${\cal A}$.
In the two-loop calculation the helicity mismatch argument operating at
one-loop is not valid any more. Thus, the resonant $ImG_{T}$ starts
contributing. So, in this calculation one hopes to compensate the suppression
from the extra powers of the coupling constant (due to the second loop)
with the resonant contributions now present, in such a way that the two-loop
resonant contributions are effectively comparable
to one-loop contributions.
In estimating ${\cal A}$ the values of
$s_{i}$,$c_{i}$, and $\delta$
have been maximized, subject to all experimental constraints.
In particular, for $M_{H^{+}}=200~GeV$, $s_{1}=0.252$,
$s_{2}= 8.29\times 10^{-3}$, $s_{3}=0.707$, and $\delta=\frac{\pi}{2}$,
we have that ${\cal A} = - 3.9\times 10^{-5}$.

\section{New one-loop contributions}

As already indicated in the previous section, there is an entire class of
graphs which contribute to ${\cal A}$ at one-loop,
which have not been
included in the original
 calculations. Such contributions originate from
imaginary parts of self-energy, vertex, and box diagrams, which contain
gauge boson loops instead of fermionic loops. The reason such graphs
contribute to ${\cal A}$ is due to the fact that ${\cal A}$ receives
contributions through the entire phase space integration range, from
$m_{\tau}^{2}$ to $m_{t}^{2}$. There are two types of such thresholds:

i) bosonic thresholds, opening when $q^{2}>M^{2}$,
($W\rightarrow W\gamma$); clearly, the imaginary
parts of such graphs contribute in the phase space
integration for $q^{2}>M^{2}$.

ii) top thresholds, corresponding to $t\rightarrow Wb$, from vertex and box
(but not $W$ self-energy) graphs.
The imaginary parts of such graphs
are non-vanishing for every value of $q^{2}$, as long as
$m_{t}^{2}>M^{2}+m_{b}^{2}$, which is of course true.

As before, only the longitudinal components contribute to ${\cal A}$
at one-loop. Moreover, such contributions are non-resonant, just as the
longitudinal $W$ self-energy graphs containing fermion loops.
However, since
there is no suppression factor $\frac{m_{c}^{2}}{M^{2}}$ in this case,
such graphs are in general expected to contribute significantly; as we will
see shortly, this is indeed the case.

Having realized the relevance of the new thresholds, their computation is in
principle straightforward. All one needs to do is isolate the longitudinal
contributions and then compute their imaginary parts. It turns out that
the process of isolating the longitudinal parts is significantly
facilitated if one uses a particular type of gauges. So, instead of
using the common choice of the renormalizable $R_{\xi}$ gauges,
we will work in the context of the background field gauges (BFG) \cite{Abb},
using appropriate Feynman rules.
The reason for this choice is the fact
that in the BFG framework, the self-energy and vertices satisfy the
following set of naive, QED-like Ward identities:
\begin{equation}
q^{\mu}q^{\nu}{\hat{\Pi}}_{\mu\nu}
-2Mq^{\mu}{\hat{\Theta}}_{\mu}+M^{2}\hat{\Omega}=0 ~~,
\label{WIa}
\end{equation}
\begin{equation}
q^{\mu}{\hat{\Pi}}^{\mu\nu}-M{\hat{\Theta}}_{\nu}=0 ~~,
\label{WIb}
\end{equation}
\begin{equation}
q^{\mu}{\hat{\Gamma}}_{\mu}-M\hat{\Lambda}=0 ~~.
\label{WIc}
\end{equation}
where
${\hat{\Pi}}_{\mu\nu}$ is the $W^{+}W^{-}$ self-energy,
${\hat{\Theta}}_{\mu}$ is the $\phi^{+}W^{-}$ mixing term,
$\hat{\Omega}$ the $\phi^{+}\phi^{-}$ self-energy,
${\hat{\Gamma}}_{\mu}$ is the $Wtb$ (or $W\tau\nu$) vertex
and $\hat{\Lambda}$ is the $\phi tb$ (or $\phi\tau\nu$) vertex,
all of them computed to one-loop, in the context of the BFG.~
$\phi^{+}$ is the charged would-be Goldstone boson.
All the above quantities depend in general on the
gauge-fixing parameter $\xi_{Q}$, used to gauge-fix the quantum
field inside the loops. However, since the final answer is guaranteed to be
$\xi_{Q}$-independent, provided {\sl all} graphs are included, any choice
for  $\xi_{Q}$ is legitimate; in particular, we choose  $\xi_{Q}=1$.

Returning to the Ward identities,
it is relatively straightforward to
exploit them, in order to decompose the amplitude in transverse and
longitudinal pieces, without detailed knowledge of the explicit closed
expressions of the individual graphs \cite{Pap}.
We define $\Gamma_{0}^{\mu}=
\frac{g}{2\sqrt{2}}\gamma^{\mu}(1-\gamma_{5})$ and
$\Lambda_{0}=
\frac{g}{2M\sqrt{2}}[m_{1}(1-\gamma_{5})-m_{2}(1+\gamma_{5})]$;
when sandwiched between on shell external spinors
$u_{1}(p_{1})$ and $u_{2}(p_{2})$, with $q=p_{1}-p_{2}$ the identity
${\bar{u}}_{1}q_{\mu}\Gamma_{0}^{\mu} u_{2}=
{\bar{u}}_{1}M_{w}\Lambda_{0} u_{2}$ holds.
Furthermore,
we define
\begin{equation}
{\hat{\Gamma}}_{\mu}^{t}
= {\hat{\Gamma}}_{\mu}+ \frac{q_{\mu}}{q^{2}}M\hat{\Lambda}~,
\label{Gt}
\end{equation}
and
\begin{equation}
{\hat{\Pi}}_{\mu\nu}^{t}=
{\hat{\Pi}}_{\mu\nu}-
\frac{q_{\mu}q_{\nu}}{q^{2}}M\hat{\Theta}~.
\label{pit}
\end{equation}
Both ${\hat{\Gamma}}_{\mu}^{t}$ are
${\hat{\Pi}}_{\mu\nu}^{t}$ are transverse, e.g.
\begin{equation}
q^{\mu}{\hat{\Gamma}}_{\mu}^{t}=0, ~~~q^{\mu}{\hat{\Pi}}_{\mu\nu}^{t}=0~~.
\label{qw}
\end{equation}
Using the identity
\begin{equation}
\frac{1}{M^{2}}= \frac{1}{q^{2}}+ \frac{q^{2}-M^{2}}{q^{2}M^{2}}~,
\label{Ident}
\end{equation}
we obtain \cite{Pap} for the propagator-like contribution
${T}_{1}$ of the
$S$-matrix element
\begin{equation}
{T}_{1}=
{\Gamma}^{\mu}_{0}[\frac{1}{q^{2}-M^{2}}]{\hat{\Pi}}_{\mu\nu}^{t}
[\frac{1}{q^{2}-M^{2}}]{\Gamma}^{\nu}_{0}+
\Lambda_{0}[\frac{1}{q^{2}}]\hat{\Omega}[\frac{1}{q^{2}}]\Lambda_{0}~,
\label{T1}
\end{equation}
and for the vertex-like piece ${T}_{2}$
\begin{equation}
{T}_{2}=
\Gamma^{\sigma}_{0}
[\frac{g_{\sigma}^{\mu}}{q^{2}-M^{2}}]
{\hat{\Gamma}}_{\mu}^{t}
-\Lambda_{0}[\frac{1}{q^{2}}]\hat{\Lambda}~.
\label{T2}
\end{equation}
It is important to notice that the longitudinal parts of
Eq(\ref{T1}) and Eq(\ref{T2})
are multiplied by the kinematic factor $\frac{1}{q^{2}}$, instead of
$\frac{1}{q^{2}-M^{2}}$; they are therefore manifestly non-resonant,
in the entire range of
the phase space integration, even at $q^{2}=M^{2}$.

\section {Calculations and results}

By virtue of this decomposition, we only need to calculate self-energy and
vertex graphs with a charged $\phi$ (but not $W$) coming in;
this represents
a significant calculational
simplification. On the other hand, since no such simple
decomposition exists for
box-like parts of the $S$-matrix,
we will compute the imaginary contributions of box diagrams
directly, and then isolate their longitudinal parts.
It turns out that graphs containing a $Z$ or a $\phi_{z}$
inside their loops
are numerically
suppressed. Since all such graphs form a gauge-invariant subset, their
omission does not interfere with the gauge independence of the final
answer.

The effect of these contributions
is additionally enhanced due to the presence of large logarithms
of the form $ln(\frac{m_{t}^{2}}{m_{b}^{2}})$,
$ln(\frac{m_{t}^{2}}{m_{\tau}^{2}})$, and
$ln(\frac{m_{t}^{2}}{M^{2}})$, which originate from vertex and box
diagrams.
After collecting all contributions and integrating over the available
phase space, using the same values for the constants
$s_{i}$, $c_{i}$ and $\delta$ and $M_{H^{+}}$ as before,
we finally find
${\cal A} = - 2.0\times 10^{-5}$.

We notice that:

i) The result of these new threshold is comparable to the outcome of the
two loop resonant calculation, and at least two
orders of magnitude larger then
the one-loop fermionic contributions.

ii) The new result comes with the same relative sign as the two-loop
result; therefore, the entire effect is to further enhance
the value of the PRA.

\section {Conclusions}

In this paper we addressed issues related to the calculation of
the PRA in the WM. We focused on semileptonic top decays,
on the dominant channel $t\rightarrow b\tau\nu$.
We showed that due to the fact that the PRA receives contributions from
the entire kinematically available phase space, new one loop contributions,
not previously considered, arise.
Such contributions are non-resonant and gauge-invariant.
 It turns out that the PRA so obtained is
two orders of magnitude larger than the one calculated form the
non-resonant fermionic contributions to the $W$ self-energy,
and are comparable to the two loop result.

\section {Acknowledgment}
The author thanks Professor
Albero Sirlin for suggesting this problem to him.
This work was supported by the National Science Foundation under
Grant No.PHY-9017585.


\end{document}